# Kinetic Monte Carlo simulation of the nitridation of the GaAs (100) surfaces

A. P. Castro and H. W. Leite Alves[†]

*Departamento de Ciências Naturais, Universidade Federal de São João del Rei, Caixa Postal 110, 36300-000 São João del Rei-MG, Brazil*

We present, in this work, our preliminary results of a systematic theoretical study of the adsorption of N over As-terminated GaAs (100) (2×1) surfaces. We analyzed the changes in the bond-lenghts, bond-angles and the energetics involved before and after deposition. Our results show that the N-atoms will prefer the unoccupied sites of the surface, close to the As dimer. The presence of the N pushes the As dimer out of the surface, leading to the anion exchange between the N and As atoms. Based on our results, we discussed about the kinetics of the N islands formation during epitaxial growth of the III-Nitrides.

## I. Introduction

The group III-nitrides (AlN, GaN, InN) and the corresponding alloys, due to their succesful applications in the electronic and optoelectronic device technology, have attracted great interest in the last 10 years. It is well known that the most stable structure for these compounds is the wurtzite one, and the optoelectronic devices based on this structure work very well, despite both the observed large piezoelectric charges induced by polarization effects parallel to their [0001] direction, and the large density of structural defects close to the substrate interface. So, for good electronic devices, it is desired to avoid this polarization effects, which is not observed in cubic structures, as well as to control the density of these structural defects [1].

However, the growth of III-nitrides in the zinc-blende structure has been a hard task to the experimentalists, once that it can not be achieved by equilibrium process techniques [2]. By using epitaxial techniques, a lot of substrates has been proposed for the III-nitride growth in the cubic modification, and the nitridation, with or without the use of As or In surfactants, of both SiC and GaAs surfaces seems to be the most efficient one for such growth [3-6].

In order to understand the physical mechanisms for III-nitride growth in the zinc-blende structure, we present, in this work, our preliminary results from a theoretical study of the adsorption of N over As-rich GaAs (100) (2×1) surfaces, based both on accurate, parameter-free, self-consistent total energy and force calculations using the density functional theory, the local-density approximation for the exchange-correlation term, within the plane-wave pseudopotential method (ABINIT code [7]), together with a kinetic Monte Carlo simulation of the epitaxial growth, based on the solid-on-solid approach [8]. We have used the Troullier-Martins pseudopotentials [9], and we have included Ga-3d electrons as valence states.

In all calculations, the slab supercells were build up of 5 atomic layers and a vacuum region equivalent of 5 atomic layers, and also, a (4 4 2) Monkhorst-Pack mesh was used to sample the surface Brillouin zone. We have followed exactly as done in our previous work on GaAs (110) surfaces [10]. The calculated total energies were used as transition rates for the kinetic processes in our Monte Carlo simulation. We have, then, analyzed the changes in the bond-lengths and in the bond-angles before and after deposition.

It is well known that the As-terminated GaAs (100) surfaces are stable, at room temperature, in a β2 (2×4) reconstruction pattern [11]. So, the (2×1) surfaces were chosen for our simulations because, at the conditions of the GaN growth on GaAs substrates, i.e., at 700º C, the GaAs (100) surface makes a phase transition to this reconstruction model [12]. Details about the structural properties of the GaAs (100) (2×1) surfaces are described in our previous work [13].

[†] Corresponding author: e-mail: hwlalves@ufsj.edu.br, tel.: +55 32 33.79.24.89, fax: +55 32 33.79.24.83

## II. N adsorption and diffusion

In Fig. 2, we show our preliminary results for the N adsorption for the sites described in Fig. 1. It is clear that the N-atoms will prefer the sites C, D and E, respectively, which are the unoccupied sites of the surface, before the surfactant effect between the N and As atoms occurs [4].

It is interesting to note that the adsorption over the As dimer (at the A site) is 1.48 eV less stable than that over the C site, which has an adsorption energy of 3.17 eV. Also, the adsorption at D and E sites are 1.09 and 1.12 eV less stable than at the C site, respectively. We noticed that, unlike what it was observed for the N adsorption over GaAs (110) surfaces [10], there is no activation barrier for N before the beginning of the adsorption process.

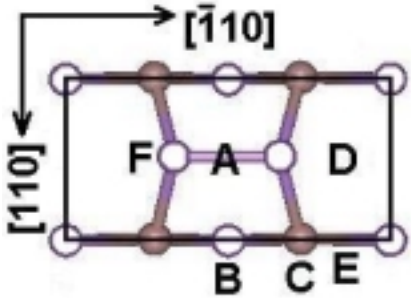

Fig. 1. Top view of the possible N adsorption sites over As-terminated GaAs (100) (2×1) surfaces. The open circles are the As atoms, while the full circles are the Ga ones.

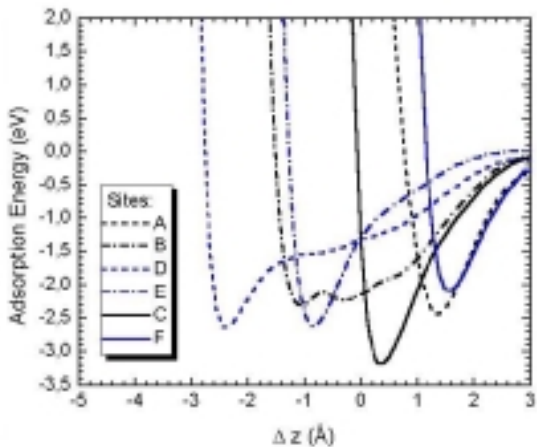

Fig. 2. Adsorption energy variation as a function of the distance of the adsorbed N atom from the As dimers, $\Delta z$ (in Å), over the GaAs (100) (2×1) surface at the sites A, B, C, D, E and F, as defined in Fig. 1.

In Table 1, we display the structural parameters for relaxation of the system after the N adsorption at C, D and E, as defined in Fig. 3. From the obtained results, we note that, only at the E site, there was a slight distortion on the $\alpha$ angle towards to a planar configuration of As bonds. Also, only at the C site, we detected a tilt of the As dimer, caused by the presence of N and, except for the adsorption at D site, the N prefers to stay close to the Ga atoms.

In order to understand the changes of the bond lengths and bond angles, we depict in Fig. 4, Relaxed atomic positions after the N adsorption at the C site. It interesting to observe that the adsorbed N, in order to stay close to Ga atoms, distorts the As dimers, pushing them away from the surface. Based on our results, we suppose that, if another N atom comes, the anion exchange between As and N atoms begins, culminating with a $As_2$ desorption. This is a possible model for the surfactant effect observed in the III-Nitride growth [4]. Calculations to check this model are currently underway.

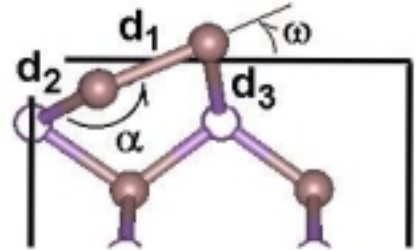

Fig. 3. Perspective view of the structural parameters which describes the atomic geometry of the GaAs (100) (2×1) surfaces. The atoms are depicted as in Fig. 1.

Table I. Calculated structural parameters, as defined in Fig. 3, for the N adsorption over GaAs (100) (2x1) surfaces at the sites C, D and E, as defined in Fig. 1.

| Parameter | C | D | E |
|---|---|---|---|
| $\alpha$ (degrees) | 102.9 | 103.6 | 112.6 |
| $\omega$ (degrees) | 11.8 | 0.0 | 0.0 |
| $d_1$ (Å) | 2.44 | 2.39 | 2.38 |
| $d_2$ (Å) | 2.88 | 2.34 | 2.30 |
| $d_3$ (Å) | 2.31 | 2.34 | 2.30 |
| $d_{N-Ga}$ (Å) | 1.64 | 2.98 | 1.70 |
| $d_{N-As}$ (Å) | 2.03 | 1.90 | 1.83 |

The energies involved in the nitridation process are relatively higher if we compare to those of the GaAs homoepitaxy, which have values at the order of 2 eV of magnitude [8,14]. During the nitridation, other processes, such as diffusion of N over the surface, also takes place. In Fig. 5, as an example, we have plotted the total energy variation as a function of the N displacement (the surface atoms were kept fixed at their relaxed positions), starting from the E adsorption site, along the [-1 1 0] direction, simulating the N diffusion along this direction. Here, the obtained energies were referenced to the calculated total energy of the N at site C. We have noted that, for a complete diffusion along the [-1 1 0] direction, the N may overcome an energy barrier of 7.34 eV. Our results have also shown that, if the diffusion takes place along the [1 1 0] direction, or over the As dimer, this energy barrier would be 5.50 eV, or 60.0 eV [15], respectively. In other words, the As dimers act as N traps and, the diffusion of N over the GaAs surface will take place only at the paths CECBC (parallel to the [-1 1 0] direction) and EDE (parallel to the [1 1 0] direction), as depicted in Fig. 1. Moreover, we expect that the N atoms have low mobility, when compared to the Ga or $As_2$ diffusion over this surface, during the nitridation process, leading to rough surfaces.

## III. Kinetic Monte Carlo Simulations

In order to check how the low N mobility affects the nitridation process, we have started a kinetic Monte Carlo simulation of the N adsorption of the As-terminated GaAs (100) (2×1) surface, based on the solid-on-solid approach, as described in Refs. 8 and 14. For this preliminary simulation, we have considered only the processes (and their respective energies) of the N adsorption at site C, as well the N diffusion along the paths [-1 1 0] and [1 1 0] [16], with a common prefactor of $10^{13}$ $s^{-1}$ [8]. We have done, so, two simulations: in the first one, we have varied the N coverage form 0.01 ML to 0.5 ML, at a fixed N flux of 0.15 ML/s, and at 700 K of growth temperature and, in the second, we have varied the N flux at three growth temperatures, 600, 700 and 800 K, keeping fixed the coverage at 0.5 ML. These results are depicted in Figs. 6 and 7.

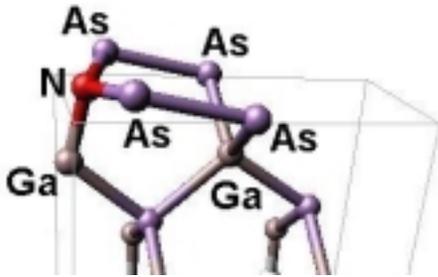

Fig. 4. Relaxed atomic positions after the N adsorption at the C site of the GaAs (100) (2×1) surface.

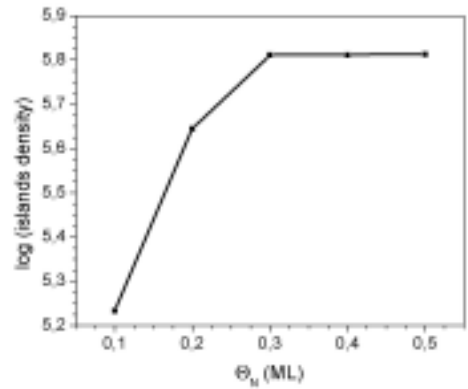

Fig. 6 Calculated variation of the logarithm of the islands density as a function of the N coverage (the growth temperature is 700 K).

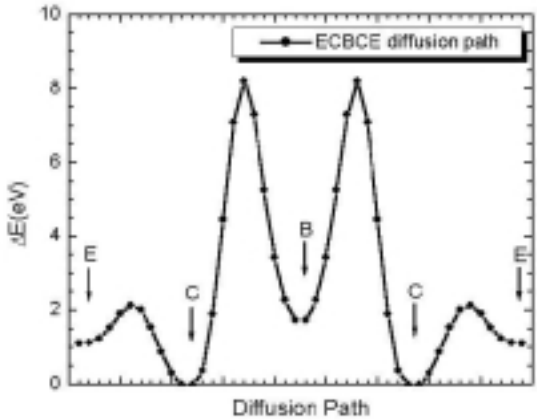

Fig. 5. Total energy variation as a function of the N displacement, starting from the E adsorption site, along the [-1 1 0] direction, simulating the N diffusion over the GaAs (100) (2×1) surface.

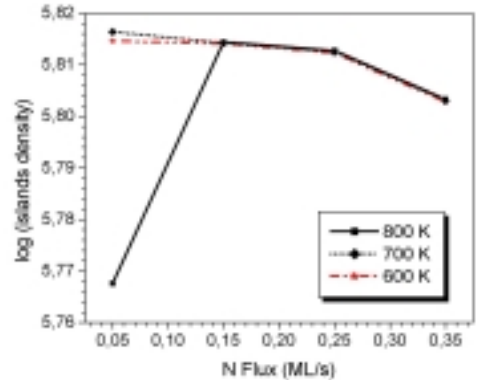

Fig. 7 Calculated variation of the logarithm of the islands density as a function of the N flux.

From Figs. 6 and 7, we have noted that, at low N flux, as well as at N coverage at the order of 0.3 ML, the density of N islands is very high. It is interesting to note that the nitridation at low N flux is efficient only at temperatures smaller than 700 K, due to the high values of islands density. Also, from our simulations, we have detected that some islands have an appreciable height of adsorbed N. This effect induces the surface roughness, and it does not diminish when we go to higher coverages – the N island density only diminish with the increasing N flux after its saturation, and this is in agreement with our obtained DFT-LDA results. Further details of our Monte Carlo simulations and a careful analysis of the obtained results will be published soon in another publication.

## IV. Final remarks

In summary, we have presented our preliminary results of a systematic theoretical study of the adsorption of N over As-terminated GaAs (100) (2×1) surfaces. We analyzed the changes in the bond-lenghts, bond-angles and the energetics involved before and after deposition. Our results show that the N-atoms will prefer the unoccupied sites of the surface, close to the As dimer. The presence of the N pushes the As dimer out of the surface, leading to the anion exchange between the N and As atoms. Also, as the N diffuses slower than As or Ga over the GaAs substrate, the nitridation induces roughness surfaces after the deposition process, which was detected by our preliminary kinetic Monte Carlo simulations of this process. We hope that our results give guidelines for future experiments on this subject.

## Acknowledgements

A. P. Castro is indebted to the Brazilian Agency, MEC-CAPES, for her fellowship during the development of this research. This work was also supported by the MCT/CNPq/PRONEX project No. 662105/98, Brazil.